\documentclass[cameraready]{Interspeech}
\usepackage{tabularx}
\usepackage{hyperref}
\usepackage{tipa}
\usepackage{xurl}
\usetikzlibrary{arrows.meta,positioning,shapes.geometric,fit,calc}
\title{HASS: Hierarchical Simulation of Logopenic Aphasic Speech for Scalable PPA Detection}

\author[affiliation={1}]{Harrison}{Li}
\author[affiliation={1}]{Kevin}{Wang}
\author[affiliation={1}]{Cheol Jun}{Cho}
\author[affiliation={1}]{Jiachen}{Lian}
\author[affiliation={2}]{Rabab}{Rangwala}
\author[affiliation={3}]{Chenxu}{Guo}
\author[affiliation={4}]{Emma}{Yang}
\author[affiliation={2}]{Lynn}{Kurteff}
\author[affiliation={2}]{Zoe}{Ezzes}
\author[affiliation={2}]{Willa}{Keegan-Rodewald}
\author[affiliation={2}]{Jet}{Vonk}
\author[affiliation={2}]{Siddarth}{Ramkrishnan}
\author[affiliation={5}]{Giada}{Antonicelli}
\author[affiliation={2}]{Zachary}{Miller}
\author[affiliation={2}]{Marilu Gorno}{Tempini}
\author[affiliation={1}]{Gopala}{Anumanchipalli}

\address{
  $^1$ UC Berkeley, USA \\
  $^2$ UCSF, USA \\
  $^3$ Zhejiang University, China \\
  $^4$ Columbia University, USA \\
  $^5$ Basque Center on Cognition, Brain and Language, Spain
}

\email{
  liharrison@berkeley.edu,
  kwang3170@berkeley.edu,
  gopala@berkeley.edu
}


\newcommand{\blue}[1]{\textcolor{blue}{#1}}
\newcommand{\red}[1]{\textcolor{red}{#1}}
\usepackage{comment}

\begin{document}

\maketitle

\begin{abstract}
    Building a diagnosis model for primary progressive aphasia (PPA) has been challenging due to the data scarcity. Collecting clinical data at scale is limited by the high vulnerability of clinical population and the high cost of expert labeling. To circumvent this, previous studies simulate dysfluent speech to generate training data. However, those approaches are not comprehensive enough to simulate PPA as holistic, multi-level phenotypes, instead relying on isolated dysfluencies. To address this, we propose a novel, clinically grounded simulation framework, Hierarchical Aphasic Speech Simulation (HASS). HASS aims to simulate behaviors of logopenic variant of PPA (lvPPA) with varying degrees of severity. To this end, semantic, phonological, and temporal deficits of lvPPA are systematically identified by clinical experts, and simulated. We demonstrate that our framework enables more accurate and generalizable detection models. Code: \url{https://github.com/haribary/HASS}

\end{abstract}

\section{Introduction}
Primary progressive aphasia (PPA) is a neurodegenerative disorder characterized by progressive language impairment. Because the hallmark of PPA is progressive deterioration of language function, connected (spontaneous) speech provides rich diagnostic information for PPA variant characterization and is widely used in clinical and research assessment \cite{wilson2018qab,matiasguiu2022spontspeech,gornotempini2011ppa}.  As speech-based machine learning methods advance, there is growing interest in automated screening frameworks for PPA \cite{rezaii2024aiclassifiesppa,peters2025interspeech, vonk2025automated}. However, such approaches are constrained by the limited availability of high-quality naturalistic speech datasets from clinically characterized PPA patients. Data collection typically requires expert diagnosis, structured elicitation protocols, and careful annotation under strict ethical and privacy constraints \cite{forbes2012aphasiabank,MacWhinney01112011,lanzi2023dementiabank}. Public resources such as DementiaBank and AphasiaBank provide invaluable connected-speech samples, but remain limited in size and institutional representation, constraining model development and cross-corpus robustness \cite{forbes2012aphasiabank,MacWhinney01112011,lanzi2023dementiabank}.

Synthetic data generation offers a potential solution, but prior dysfluency simulation efforts have largely focused on injecting isolated dysfluency events (e.g., repetitions, insertions, or pauses) into otherwise fluent speech \cite{lian2023unconstrained-udm, ssdm,zhang2025llmdys,lian2024ssdm20timeaccuratespeech,zhou2024yolostutter,kourkounakis2021fluentnet}. However, behavioral deficits in some PPA variants arise from disruptions on multiple levels (phonological, word and content) of speech \cite{gornotempini2011ppa}. Phonological deficits simulated in isolation fail to capture these interactions. As a result, such simulations often lack clinical plausibility and fail to model simultaneous disruptions across phoneme, lexical, and content levels. Furthermore, most simulations using LLM-based or agentic generation pipelines are not grounded in the production mechanisms of a specific clinical phenotypes, limiting their utility because models trained on such data may learn generic surface-level dysfluency patterns rather than disorder-specific impairment signatures \cite{zhang2025llmdys,kourkounakis2021fluentnet,imaezue2025abcd,pittman2025syntheticpwa}. To date, no end-to-end simulation framework has explicitly modeled a neurodegenerative language disorder using clinically grounded production mechanisms.

Among PPA variants, lvPPA is characterized by impaired word retrieval that cascades into phonological errors and halting speech \cite{gornotempini2011ppa,gornotempini2008logopenic}, making it an ideal test case for a multi-level simulation framework. Thus, we introduce the Hierarchical Aphasic Speech Simulation (HASS), a clinically grounded hierarchical simulation framework for logopenic variant PPA (lvPPA) that models clinically defined impairment mechanisms in a two-layer production model: (1) a lexical retrieval impairment layer that generates severity-conditioned content-level disruptions and (2) a phonological encoding disruption layer that introduces severity-conditioned phoneme-level errors on a word-aligned representation. HASS produces severity-controlled synthetic lvPPA speech with co-occurring content-level and phonological dysfluencies, alongside matched controls generated using the same pipeline but without impairment injection. This enables scalable augmentation of low-resource clinical datasets while ensuring classifier differences are attributable to the simulated disorder rather than synthesis artifacts. Our experiments show that HASS improves the performance of the diagnosis model during evaluation. We will release all data, models, and code to support reproducibility. We summarize our contributions as follows:

\begin{itemize}
\item We propose HASS, an exclusive clinician-guided simulation pipeline and the first framework to model a neurodegenerative aphasia (lvPPA) as a holistic, structural disease rather than a collection of isolated, disconnected dysfluencies.
\item We introduce a scalable recipe for clinical data augmentation, releasing a comprehensive, severity-controlled synthetic dataset that accurately reflects the multi-level impairments of PPA.
\item We demonstrate that HASS-generated data improves automated PPA classification, with classifiers trained on HASS speech outperforming those trained on clinical recordings. We evaluate both in-domain capability and cross-site generalization.
\end{itemize}

\begin{figure*}[t]
    \centering
    \includegraphics[width=\textwidth]{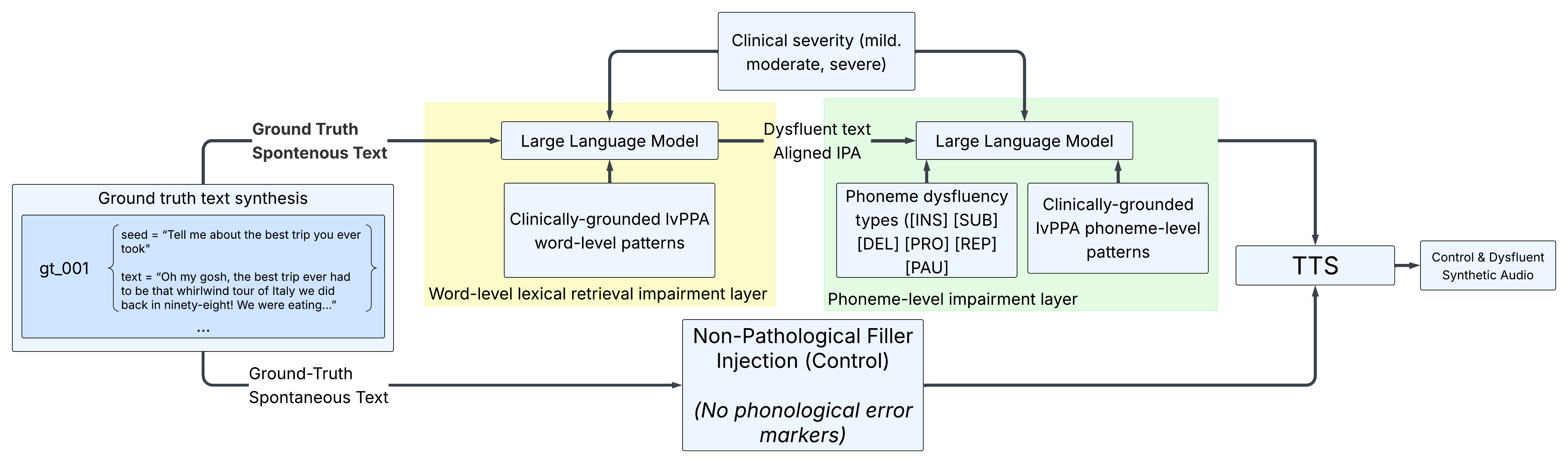}
    \caption{Overview of the HASS hierarchical simulation pipeline.}
    \label{fig:simpipeline}
\end{figure*}

\section{Simulation}

\subsection{Dysfluent Text Generation}

We introduce a two-layer dysfluent text generation pipeline designed to model language production deficits characteristic of Logopenic Variant primary progressive aphasia (lvPPA). In particular, we recruit an LLM (Gemini 3 \cite{geminiteam2023gemini}) to simulate pathological behaviors with detailed, clinically guided instructions. The simulator encodes clinically defined lvPPA symptoms, and was developed with SLP oversight. LvPPA's primary impairment is in lexical retrieval and its downstream phonological consequences, which inspired the factorization of dysfluency modeling into content-level and phoneme-level processes.

During generation, the LLM is constrained by several clinically grounded rules: \begin{itemize} \item \textbf{Lexical Bias:} In lvPPA, content and phonological dysfluencies are correlated. Dysfluencies are applied non-uniformly, heavily biasing toward high lexical-demand loci such as low-frequency content words and multisyllabic targets. \item \textbf{Syntactic Adherence:} Disruptions are constrained to plausible syntactic and discourse boundaries (e.g., clause boundaries or pre-content-word positions). \item \textbf{Phenotype Exclusion:} Features characteristic of non-fluent/agrammatic and semantic PPA variants (e.g., persistent agrammatism, apraxia-of-speech-like distortions, or semantically empty fluent speech) are explicitly penalized to prevent phenotype drift. \end{itemize}

\subsubsection{Word Level}

Non-pathological spontaneous text is first generated using a diverse set of prompts modelled in the style of connected speech questions sourced from the Quick Aphasia Battery\cite{wilson2018qab}. The output is directed into two pipelines: the synthetic control pipeline data are injected with naturalistic dysfluencies on the word level and the synthetic dysfluency pipeline where this text now serves as the ground truth.

In the dysfluency pipeline, ground-truth text is first passed to the word-level (content) dysfluency layer. Conditioned on the chosen severity variable, we instruct an LLM to introduce lvPPA-like lexical retrieval phenomena, including circumlocutions, false starts, and filled pauses, while preserving the intended message. The resulting word-level dysfluent text is then converted to a word-aligned IPA representation using \cite{Bernard2021}. Stress markers and word boundaries are retained to enable consistent alignment for downstream phonological editing and speech synthesis.

\subsubsection{Phoneme Level}

The phoneme-level layer edits the word-aligned IPA sequence, conditioned on both the word-level dysfluent text and its IPA target form. We instruct LLMs to insert inline markers for six error types organized in a clinically motivated hierarchy. The three primary markers are \texttt{[PAU]} (pause insertion), \texttt{[SUB]} (phoneme substitution), and \texttt{[DEL]} (phoneme deletion), reflecting the dominant temporal and phonological disruptions in lvPPA, where slow speaking rate, word-finding halts, and phonological paraphasias predominate \cite{aftd2018lvpaa,gornotempini2008logopenic,gornotempini2011ppa,henry2018assessment,petroi2020phonologic}. Next, two secondary markers are also modelled: \texttt{[REP]} (sound/syllable repetition), modeled as a byproduct of self-repair during failed retrieval \cite{gornotempini2011ppa}, \texttt{[PRO]} (phoneme prolongation), reflecting mild hesitation-related lengthening \cite{gornotempini2011ppa}. Finally, \texttt{[INS]} (phoneme insertion), is treated as a rare dysfluency since it is reported at lower rates than other phonological paraphasias \cite{dalton2018paraphasias}. Furthermore, marker rates are severity-conditioned and biased toward content words ($\geq$80\%), with disruption probability increasing with word length and syllable complexity. Repetition is restricted to repair contexts, and within-word marker co-occurrence is capped at higher severities to maintain a realistic density of dysfluencies. The output is a marked IPA string used for subsequent speech synthesis.

\subsection{Synthesis of Dysfluent Speech}

\begin{table*}[t]
\centering
\caption{Example output across severity levels for a single ground-truth sentence. \textbf{Text}: word-level dysfluent output. \textbf{IPA}: phoneme-level output with inline markers. Ground truth: \emph{``The house would go completely dark, save for the single amber glow of a hearth fire.''}}
\label{tab:samples}
\renewcommand{\arraystretch}{1.3}
\small
\begin{tabularx}{\textwidth}{@{}l X@{}}
\hline
\textbf{Severity} & \textbf{Output} \\
\hline
\textsc{Control} &
\textbf{Text:} The house would go completely dark, save for the single amber glow of a hearth fire. \newline
\textbf{IPA:} \textipa{D@ h"aUs wUd g"oU k@mpl"iùtli d"Aùôk, s"eIv f3ôD@ s"INg@l "\ae mb3ô gl"oU @v@ h"Aùôh f"aI@ô} \\
\hline
\textsc{Mild} &
\textbf{Text:} The house would go comple\red{\textbf{\texttt{[DEL]}}}ly dark, \blue{except for} the single, \blue{you know}, the ori\red{\textbf{\texttt{[SUB]}}}mge light of the, \blue{the place where you burn the wood}, the hearth fire. \newline
\textbf{IPA:} \textipa{D@ h"aUs wUd g\`oU k@mpl"iù}\red{\textbf{\texttt{[DEL]}}}\textipa{li d"Aùôk, Eks"Ept f3ôD@ s"INg@l, juù n"oU, DI "Oôim}\red{\textbf{\texttt{[SUB]}}}\textipa{dZ l"aIt VvD@, D@ pl"eIs w\`Eô juù b"3ùn D@ w"Ud, D@ h"Aùôh f"aI@ô} \\
\hline
\textsc{Moderate} &
\textbf{Text:} The house would go, \blue{it would go} comple\red{\textbf{\texttt{[DEL]}}}ly dark, \blue{except for the, the} oran\red{\textbf{\texttt{[SUB]}}}z light, the am\red{\textbf{\texttt{[DEL]}}}ber glow \blue{from the, the place where you burn the wood,} \red{\textbf{\texttt{[PAU]}}} the hea\red{\textbf{\texttt{[DEL]}}}th. \newline
\textbf{IPA:} \textipa{D@ h"aUs wUd g"oU, It wUd g\`oU k@mpl"iù}\red{\textbf{\texttt{[DEL]}}}\textipa{li d"Aùôk, Eks"Ept f3ôD@, DI "OôIndz}\red{\textbf{\texttt{[SUB]}}} \textipa{l"aIt, DI "\ae m}\red{\textbf{\texttt{[DEL]}}}\textipa{b3ô gl"oU fô VmD@, D@ pl"eIs w\`Eô juù b"3ùn D@ w"Ud,} \red{\textbf{\texttt{[PAU]}}} \textipa{D@ h"Aù}\red{\textbf{\texttt{[DEL]}}}\textipa{T} \\
\hline
\textsc{Severe} &
\textbf{Text:} \blue{It was, it went, uh,} \red{\textbf{\texttt{[PAU]}}} \blue{no} ligh\red{\textbf{\texttt{[DEL]}}}, \blue{just the, the one, the} fi\red{\textbf{\texttt{[PRO]}}}re thin\red{\textbf{\texttt{[SUB]}}}, \blue{in\dots}inside\red{\textbf{\texttt{[REP]}}} \newline
\textbf{IPA:} \textipa{It w"Vz, It w"Ent, "V,} \red{\textbf{\texttt{[PAU]}}} \textipa{n"oU l"aI}\red{\textbf{\texttt{[DEL]}}}\textipa{, dZ"Vst D@, D@ w"Vn, D@ f"aI}\red{\textbf{\texttt{[PRO]}}}\textipa{@ô T"In}\red{\textbf{\texttt{[SUB]}}}\textipa{, I\dots Ins"aId} \red{\textbf{\texttt{[REP]}}} \\
\hline
\end{tabularx}
\end{table*}

\begin{figure}[t]
    \centering
    \includegraphics[height=8cm]{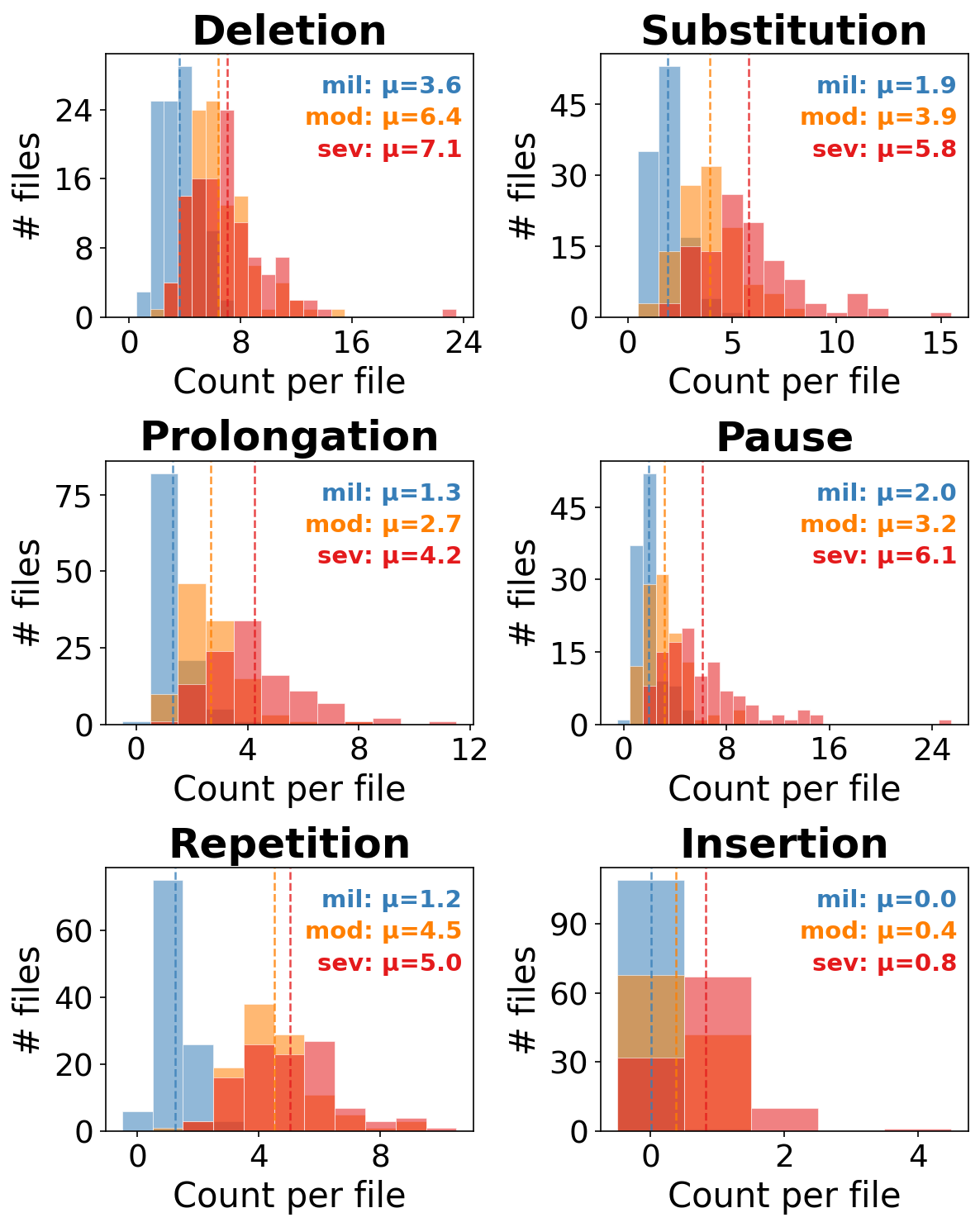}
    \caption{Distribution of phonological dysfluency markers across severity levels. Dashed lines indicate per-severity means.}
    \label{fig:marker_dist}
\end{figure}

We synthesize speech with TTS (VITS)\cite{kim2021conditionalvariationalautoencoderadversarial} while explicitly preserving dysfluency. Phoneme-level markers ([DEL], [SUB], [INS], [REP]) are applied upstream during IPA generation, while [PAU] is realized by inserting a silence segment during and [PRO] by prolonging the target phoneme during inference. We provide both sentence-level audio outputs and concatenated utterances; sentence audio is concatenated downstream using a 50\,ms crossfade.

\section{Data}

\subsection{Marker Distribution Analysis}

Figure~\ref{fig:marker_dist} confirms that the generated data respect the clinical marker hierarchy. Across all severity levels, pause, deletion, and substitution account for the majority of dysfluency events, while prolongation and repetition occur less frequently and insertion remains rare. Distributions shift toward higher counts with increasing severity: summing mean counts across markers yields $T_{\mathrm{mild}}=10.0$, $T_{\mathrm{mod}}=21.1$, and $T_{\mathrm{sev}}=29.0$ markers per file ($2.1\times$ and $2.9\times$ increases relative to mild). The three primary markers account for $75.0\%$ of mild events ($7.5/10.0$), $64.0\%$ of moderate ($13.5/21.1$), and $65.5\%$ of severe ($19.0/29.0$). At the extremes, insertion averages fewer than one event per file even at severe ($\mu_{\mathrm{INS}} = 0.0, 0.4, 0.8$), whereas the dominant markers reach substantially higher rates ($\mu_{\mathrm{DEL}}=7.1$, $\mu_{\mathrm{SUB}}=5.8$, $\mu_{\mathrm{PAU}}=6.1$ at severe).

\subsection{Dataset Composition}
The HASS corpus comprises 4{,}773 sentence-level clips totalling 12.81 hours of synthesized audio. Of these, 2{,}007 are control utterances and 2{,}766 are dysfluent (871 mild, 1{,}101 moderate, 794 severe). Speech is synthesized using 95 speakers from the VCTK corpus (out of 109 available voices) across 40 unique ground-truth prompts. Controls are generated through the same synthesis pipeline, speakers, and prompts but without lvPPA-specific dysfluency injection, ensuring that any classifier differences are attributable to the simulated impairment rather than speaker or synthesis artifacts.
\section{Experiments and Results}

\begin{table}[t]
\centering
\caption{Cross-site performance comparison (mean $\pm$ std).}
\label{tab:cross_site_results}
\begin{tabular}{lccc}
\hline
\textbf{Model} & \textbf{AUC} & \textbf{F1} & \textbf{Recall (Dys)} \\
\hline
Baseline  & 0.850 $\pm$ 0.122 & 0.778 $\pm$ 0.165 & 0.659 $\pm$ 0.238 \\
HASS  & 0.892 $\pm$ 0.076 & 0.800 $\pm$ 0.072 & 0.899 $\pm$ 0.066 \\
\hline
\end{tabular}
\end{table}

To validate the clinical utility of our synthetic corpus, we evaluate a classification model trained on HASS-generated data and assess its zero-shot generalization on real-world clinical recordings.
We evaluate on real lvPPA patient audio from the Baycrest PPA Protocol corpus \cite{baycrestcorpus} and the Hopkins PPA corpus \cite{tippett2017ppa} in DementiaBank, and on two control datasets: the Delaware corpus \cite{lanzi2023dementiabank} from DementiaBank and the Capilouto corpus \cite{capiloutocorpus} from AphasiaBank. Both control datasets are selected using explicit exclusion criteria (e.g., no neurological or cognitively deteriorating conditions, fluent English, and no clinically significant depression). Baycrest, Delaware, and Capilouto share the standard TalkBank discourse protocol~\cite{forbes2012aphasiabank,MacWhinney01112011,lanzi2023dementiabank}, whereas Hopkins uses a clinical assessment battery~\cite{tippett2017ppa} comprising naming, passage reading, counting, and story retelling. Classifiers are evaluated in a strictly cross-site design: trained on Baycrest lvPPA and Delaware controls and tested on Hopkins dysfluent and Capilouto controls, then vice versa. This protocol and recording mismatch across four independent sites provides a stringent test of cross-corpus generalization. 

\subsection{Modeling details}
\begin{figure}[h]
    \centering
    \includegraphics[width=0.75\linewidth]{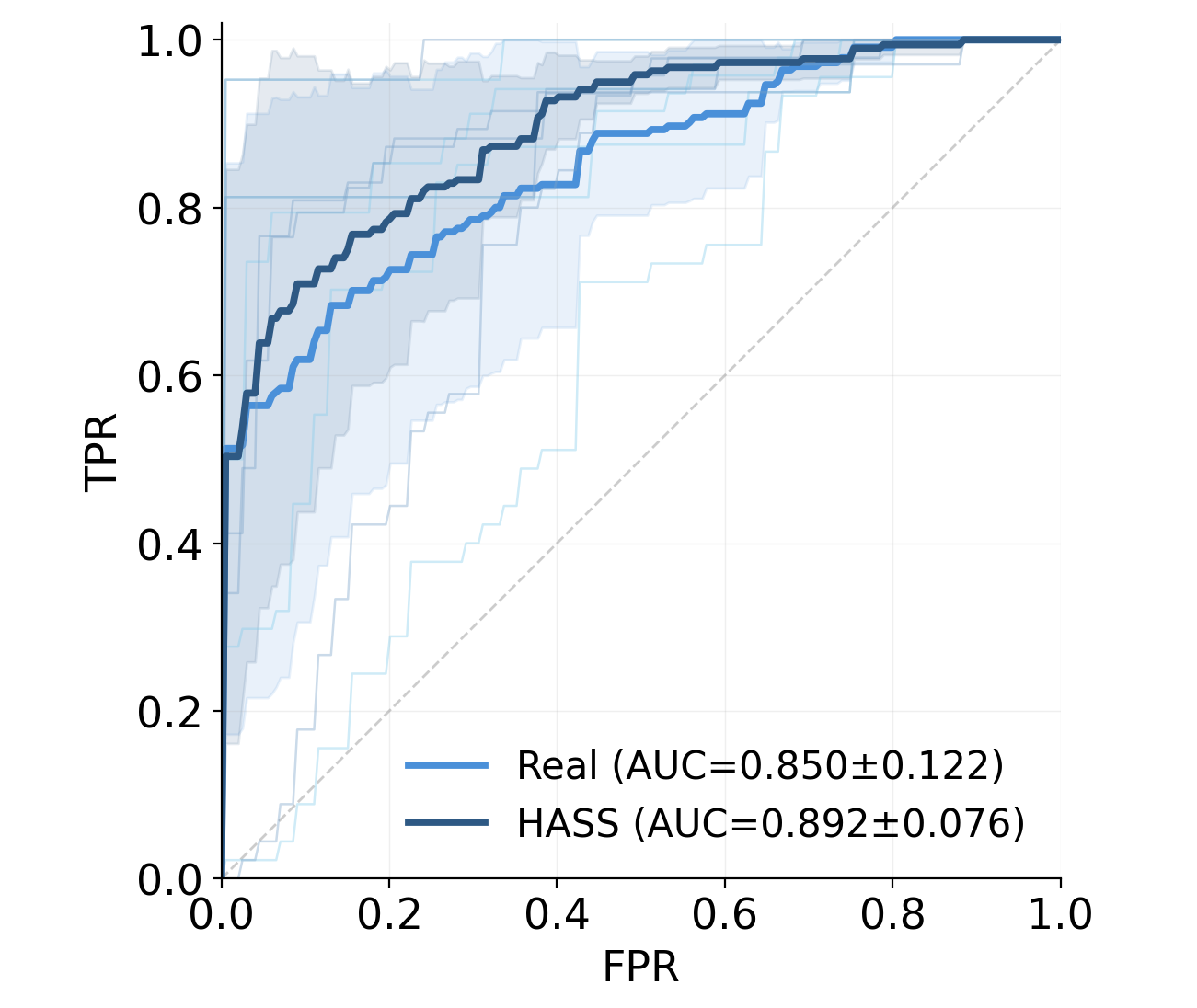}
    \caption{Comparison of ROC curves for LoRA models trained on real vs HASS speech, evaluated using 5-fold cross-validation. Mean ROC curve and ±1 standard deviation shading.
    }
    \label{fig:lora_compare_folds}
\end{figure}
We fine-tune Wav2Vec 2.0 \cite{baevski2020wav2vec} with the base model size using Low-Rank Adaptation (LoRA) \cite{hu2022lora}. We apply LoRA adapters exclusively to the query and value projection layers ($Q_{\mathrm{proj}}$, $V_{\mathrm{proj}}$). We compare two distinct training regimes:
\begin{itemize}
    \item \textbf{Baseline}: Due to the limited size of the real clinical datasets, we employ 5-fold cross-validation to ensure reliable performance estimates. Data is grouped by speaker to prevent train/test contamination. To mitigate confounding background variability, all baseline train data is enhanced with MossFormer2\_SE\_48K \cite{zhao2024mossformer2combiningtransformerrnnfree} prior to feature extraction.
    \item \textbf{HASS}: This model is trained using all samples from our generated HASS corpus. We sample fixed-length 15\,s windows (240{,}000 samples at 16\,kHz) from concatenated synthetic audio, which is grouped by ground-truth prompt, severity, and speaker. 
\end{itemize}
We apply on-the-fly augmentation with speed perturbation (0.9--1.1), additive Gaussian noise (10--20\,dB SNR), volume jitter (\(\pm 6\)\,dB), and reverberation via synthetic room impulse responses (RT60 0.2--0.8\,s), with per-sample probabilities of 0.5, 0.5, 0.5, and 0.3, respectively.
Each test fold is balanced to the minority class and evaluated on \emph{pre-enhanced} audio. We report AUC-ROC, macro F1, and recall metrics all five folds.

\subsection{Cross-Site Model Evaluation}
\begin{figure}[t]
    \centering
    \includegraphics[width=\linewidth]{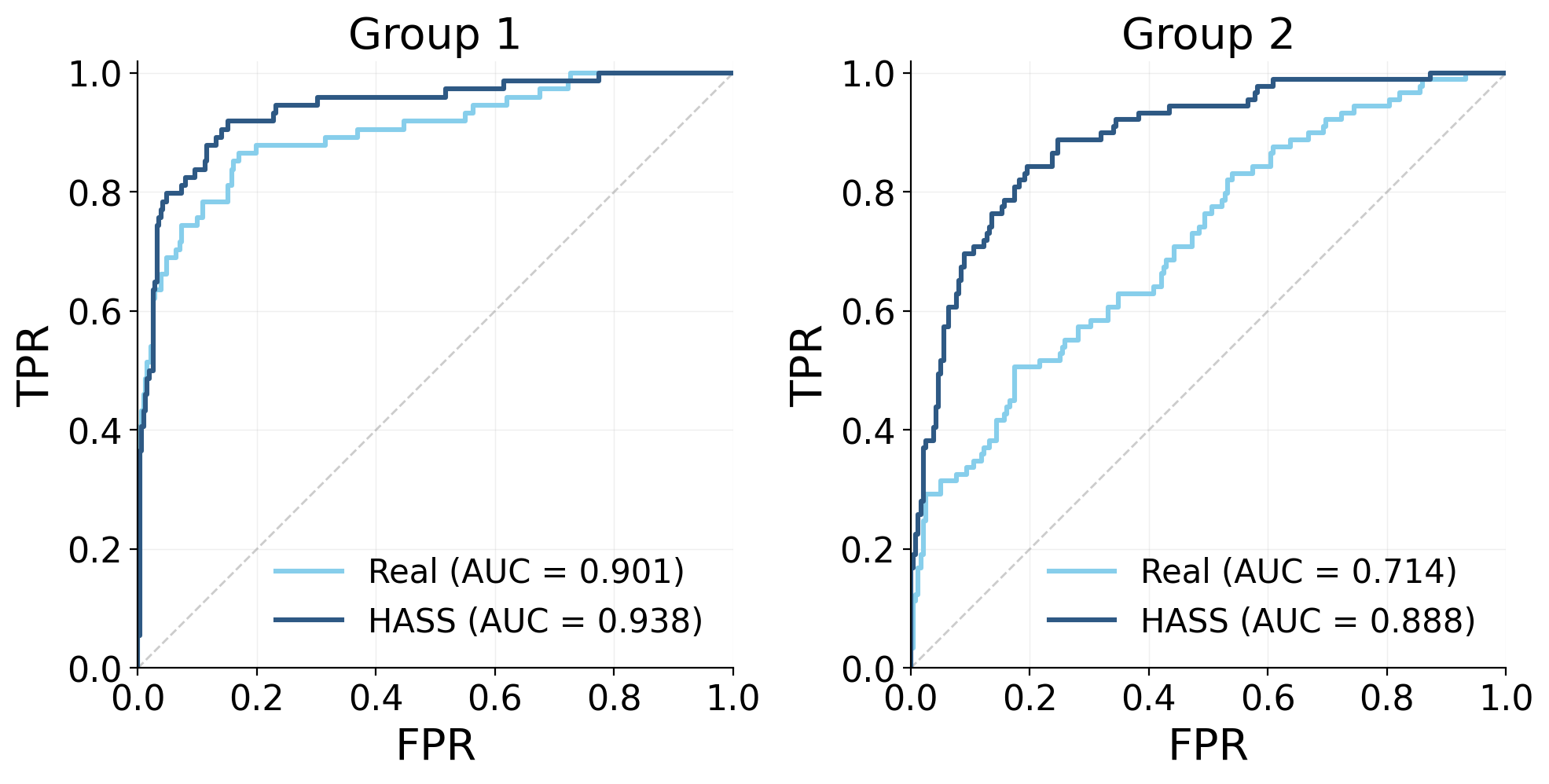}
    \caption{Cross-site ROC curves for LoRA SFT on w2v2-base. Test data as follows - 
    Group 1: JHU + Capilouto. Group 2: Baycrest + Delaware.}
    \label{fig:real_vs_hass_roc}
\end{figure}

To further evaluate the advantage of our simulated data, we compare models trained on synthetic speech against baseline models in a strict cross-site scenario. In particular, we evaluate generalization by training a model on data from one clinical site and testing it on another. We partition our real datasets into two evaluation domains: Domain A comprises Baycrest (dysfluent) and Delaware (control), while Domain B comprises JHU (dysfluent) and Capilouto (control). We separate Baycrest and JHU into different domains due to their contrasting protocols for cross-domain evaluation. All models are trained with the same setting as \S 4.1 using Wav2Vec 2.0 base.

\begin{itemize}
\item \textbf{Baseline}: Trained on real patient recordings and controls. We train a separate baseline model for each cross-site scenario (i.e., trained on Domain A and tested on Domain B, and vice versa).
\item \textbf{HASS}: A single model is trained exclusively on HASS-generated streams. This model is then evaluated on the real cross-site test domains.
\end{itemize}

\subsection{Results} 

\textbf{HASS provides synthetic samples that enable more accurate automatic lvPPA diagnosis} Table~\ref{tab:cross_site_results} summarizes performance of the comparison of HASS-trained model and the baseline models  The HASS model outperforms the baseline models which only use limited real-world data across all primary metrics. Figure~\ref{fig:lora_compare_folds} illustrates this advantage, as a HASS-trained model not only achieves higher mean AUC-ROC, but displays tighter variance across folds, indicating a more stable learning signal.

\textbf{HASS-trained models demonstrate robust cross-site generalization.} We evaluated cross-site classification capability of HASS-trained model. As illustrated in Figure~\ref{fig:real_vs_hass_roc}, the model trained exclusively on HASS-generated data outperforms baseline models trained on real clinical recordings when evaluated on recordings from different clinical sites. Notably, the HASS-trained model achieves a higher AUC-ROC in a strict cross-site setting (trained on one institutional dataset and tested on another, and vice versa). 
Cross-site robustness is essential for real-world clinical utility. Overfitting to local variations in recording environments and elicitation protocols remains the primary barrier to deploying universal, scalable diagnostic models.
This highlights two key advantages of our synthetic approach: generalization and scalability. HASS provides a mechanism to generate virtually limitless, highly diverse training samples that capture the core clinical phenotype of lvPPA without overfitting to the acoustic artifacts or demographic biases of a single clinical site.

\section{Conclusion}
In this work, we introduced HASS, a novel, clinician-guided hierarchical simulation framework designed to address the critical data scarcity bottleneck in automated primary progressive aphasia (PPA) screening. By explicitly modeling the complex, multi-level language deficits characteristic of the logopenic variant (lvPPA) at both the lexical and phonological levels, HASS generates highly realistic, severity-controlled synthetic speech. Our empirical evaluations demonstrate that the diagnostic model trained on HASS-generated data not only outperform baselines trained strictly on real clinical recordings, but also exhibit superior robustness and generalization in stringent, cross-site evaluations. Ultimately, this framework provides a scalable, privacy-preserving pathway for augmenting low-resource clinical datasets, paving the way for more reliable and robust speech-based diagnostic tools for neurodegenerative diseases.

\textbf{Limitations.} Standard phoneme-to-speech architectures like VITS are inherently optimized for fluent speech and may not perform when forced to generate severe phonological errors. Furthermore, the neurological variability of dysfluent speech remains highly complex and under active study; because individual symptoms progress heterogeneously, true PPA severity manifests in more nuanced ways.

\bibliographystyle{IEEEtran}
\bibliography{mybib}

\end{document}